\documentclass[twocolumn,tighten,numberedappendix]{aastex63}
\usepackage{amsmath}
\usepackage{ulem}
\usepackage{stmaryrd}
\setcitestyle{notesep={ }}

\graphicspath{{./}{figures/}}

\newcommand{\alfven}{{Alfv\'{e}n}}

\def\simlt{\lower.5ex\hbox{$\; \buildrel < \over \sim \;$}}
\def\simgt{\lower.5ex\hbox{$\; \buildrel > \over \sim \;$}}

\def\beq{\begin{equation}}
\def\eeq{\end{equation}}
\def\ba{\begin{eqnarray}}
\def\ea{\end{eqnarray}}

\definecolor{ao(english)}{rgb}{0.0, 0.5, 0.0}

\begin{document}

\title{
Propagation of a Strong Fast Magnetosonic Wave in the Magnetosphere of a Neutron Star
}

\author[0000-0002-4738-1168]{Alexander Y. Chen}
\affil{Physics Department and McDonnell Center for the Space Sciences, Washington University in St. Louis; MO, 63130, USA}

\author[0000-0002-0108-4774]{Yajie Yuan}
\affiliation{Physics Department and McDonnell Center for the Space Sciences, Washington University in St. Louis; MO, 63130, USA}

\author[0000-0003-0750-3543]{Xinyu Li}
\affil{Canadian Institute for Theoretical Astrophysics, 60 St George St, Toronto, ON M5R 2M8}

\author{Jens F. Mahlmann}
\affil{Department of Astrophysical Sciences, Princeton University, Princeton, NJ 08544, USA}

\correspondingauthor{Alex Chen}
\email{cyuran@wustl.edu}

\begin{abstract}
    We study the propagation of a strong, low frequency, linearly polarized fast
    magnetosonic wave inside the magnetosphere of a neutron star. The relative
    strength $\delta B/B$ of the wave grows as a function of radius before it
    reaches the light cylinder, and what starts as a small perturbation can grow
    to become nonlinear before it escapes the magnetosphere. Using
    first-principles Particle-in-Cell (PIC) simulations, we study in detail the
    evolution of the wave as it becomes nonlinear. We find that an initially
    sinusoidal wave becomes strongly distorted as $\delta B/B$ approaches order
    unity. The wave steepens into a shock in each wavelength. The plasma
    particles drift into the shock and undergo coherent gyration in the rest of
    the wave, and subsequently become thermalized. This process quickly
    dissipates the energy of an FRB emitted deep within the magnetosphere of
    magnetar, effectively preventing GHz waves produced in the closed field line
    zone from escaping. This mechanism may also provide an effective way to launch shocks
    in the magnetosphere from kHz fast magnetosonic waves without requiring a relativistic ejecta. The resulting shock can propagate
    to large distances and may produce FRBs as a synchrotron maser.
\end{abstract}



\section{Introduction}

Neutron stars are capable of producing low frequency fast magnetosonic (fms) waves
from within the magnetosphere. Star quakes in the crust can launch \alfven{}
waves along the magnetic field \citep[e.g.][]{2020ApJ...897..173B}, which can
convert spontaneously to fms waves due to propagation on curved field lines
\citep{2021ApJ...908..176Y}. These waves have a low characteristic frequency of
$\omega\sim\,10\,\mathrm{kHz}$ \citep{1989ApJ...343..839B}, much lower than the
plasma and gyro-frequencies within the magneosphere.

Fast radio bursts (FRBs), in some theoretical models, are also believed to
originate from within the magnetosphere of neutron stars (in particular
magnetars) initially as fast magnetosonic waves. These bursts are short, millisecond
duration signals in the MHz to GHz radio frequencies \citep[see e.g.][for a review]{2019A&ARv..27....4P}, and recent observations of
a simultaneous FRB and an X-ray burst from a galactic magnetar SGR 1935+2154
have corroborated the association of FRBs with strongly magnetized neutron stars
\citep[see e.g.][]{2020Natur.587...54C,2020Natur.587...59B,2020ApJ...898L..29M}.
Two broad classes of theoretical models for FRBs exist. One invokes a shock
produced by a relativistic ejecta in the wind at large distances, and both
theory and simulations have shown that the shock front is capable of generating
coherent GHz emission through the synchrotron maser mechanism
\citep{2014MNRAS.442L...9L, 2019MNRAS.485.3816P, 2021PhRvL.127c5101S}. The other
class of models invoke plasma instabilities close to the magnetar, either
creating ``bunches'' of charged particles which radiate coherently
\citep{2017MNRAS.468.2726K,2020MNRAS.498.1397L} or producing GHz waves in
reconnecting current sheets \citep{2020ApJ...897....1L}. If the FRB is produced
close enough to the star, then its frequency may be lower than the local plasma
and gyro-frequencies, and its behavior will be similar to the low frequency fms
waves produced by star quakes.

As these low frequency fms waves propagate outwards, they may become nonlinear
($\delta B\gtrsim B_\mathrm{bg}$), and it was recently proposed that these waves
may suffer very strong scattering \citep{2022PhRvL.128y5003B,
  2021ApJ...922L...7B} by charged particles in the magnetosphere, depending on
their propagation angle with respect to the background magnetic field
\citep{2022MNRAS.515.2020Q}. However, this scenario was studied only considering
the response of a single test particle. If the wave becomes nonlinear while
collective plasma effects are important (e.g. when plasma frequency exceeds the
wave frequency, $\omega_{p} > \omega$), then the plasma response may
significantly alter the wave itself, potentially leading to a completely
different dissipation mechanism. In this paper, we use first-principles
Particle-in-Cell (PIC) simulations to study the evolution of the low frequency
fms wave as it becomes nonlinear in a dense plasma. We will consider both
${\sim}\mathrm{kHz}$ waves produced by seismic activity on the star, as well as
an FRB produced deep within the magnetosphere. We will discuss how the results
apply to each case, and potential observational implications.

\section{Propagation of Fast Magnetosonic Waves}
\label{sec:fms-waves}

The highly magnetized surroundings of a magnetar is often considered to be
approximately ``force-free'', where the plasma inertia is negligible compared to
the magnetic field energy density. This is the limit of magnetohydrodynamics
(MHD) where the eletromagnetic force on the plasma is zero,
$\rho \boldsymbol{E} + \boldsymbol{j}\times \boldsymbol{B} = 0$. Two assumptions
are required for the force-free approximation:
$\boldsymbol{E}\cdot \boldsymbol{B} = 0$, and $E < B$. Under this approximation,
MHD waves simplify to only two modes: the \alfven{} mode, which has its electric
field vector $\delta\boldsymbol{E}$ in the plane of the wave vector
$\boldsymbol{k}$ and the background magnetic field $\boldsymbol{B}_\mathrm{bg}$;
and the fast magnetosonic (fms) mode, which has $\delta\boldsymbol{E}$
perpendicular to the $\boldsymbol{k}$-$\boldsymbol{B}_\mathrm{bg}$ plane
\citep[see e.g.][]{1986ApJ...302..120A}. The \alfven{} mode propagates along the
background magnetic field, while only the fms mode can propagate freely across
the field lines and escape the magnetosphere. Therefore, in this paper, we only
consider fast modes that are produced in the closed zone and propagate
quasi-perpendicularly to the background magnetic field.

In the closed zone of the neutron star, where the background magnetic field
scales as $B_\mathrm{bg}\sim r^{-3}$, the amplitude of the fast wave scales as
$\delta B\sim r^{-1}$ since it expands spherically from the emission site.
Therefore, the relative amplitude of the wave increases rapidly as
$\delta B/B_\mathrm{bg}\sim r^{2}$. Consider a cosmological FRB produced deep
within the magnetosphere as a coherent fms wave. Its luminosity determines the
radius $r_\mathrm{NL}$ where its amplitude becomes comparable to the background
magnetic field:
\begin{equation}
    \label{eq:rnl}
    r_\mathrm{NL} \sim \sqrt{B_{0}r_{*}^{3}}\left(\frac{L}{c}\right)^{-1/4}\sim 1.3\times 10^{8}\,B_{0,14}^{1/2}L_{42}^{-1/4}\,\mathrm{cm}
\end{equation}
For all known magnetars, this is far within the light cylinder,
$r_\mathrm{NL} \ll r_\mathrm{LC}$. Take the galactic magnetar SGR 1935+2154 as
an example: its spin period is $P\approx 3.2\,\mathrm{s}$ and its spindown
magnetic field is $B_{0}\sim 2\times 10^{14}\,\mathrm{G}$. An FRB of isotropic
equivalent luminosity $L \gtrsim 2\times 10^{34}\,\mathrm{erg/s}$ will have a
nonlinear radius $r_\mathrm{NL}$ within the light cylinder. The two consecutive
radio bursts observed in 2020 satisfy this criterion since they have isotropic
luminosities of order $L\sim 10^{37}\,\mathrm{erg/s}$ \citep{2020Natur.587...54C}.

Consider the simplified case where the wave propagates perpendicularly to the
background field $\boldsymbol{B}_\mathrm{bg}$. The wave magnetic field
$\delta\boldsymbol{B}$ is perpendicular to the wave vector $\boldsymbol{k}$ and
the wave electric field $\delta\boldsymbol{E}$, therefore it lies along the
background magnetic field. When $\delta B/B_\mathrm{bg} > 1/2$, the combined
magnetic field $B_\mathrm{tot}$ may become smaller than the wave electric field.
This can potentially create local $E > B$ regions which breaks ideal MHD, which
is also the basis of the force-free approximation. If the plasma remains well
magnetized, $\omega_{B} \gg \omega_{p} \gg \omega$, then its response may
prevent such $E > B$ regions to develop in the first place. The plasma response
involves its inertia in nature and cannot be probed through force-free
calculations. In a typical force free simulation, in order to preserve the
force-free condition, any $E > B$ is brought to $E = B$, resulting in artificial
dissipation \citep[e.g.][]{2019ApJ...881...13L}.

Whether plasma remains well magnetized at the nonlinear radius $r_\mathrm{NL}$
depends on its density distribution over radius $n_{\pm}(r)$. Magnetars are
known to emit X-rays that are often more luminous than their spindown power,
therefore some form of magnetic dissipation is in general believed to occur in
the magnetosphere \citep[e.g.][]{1996ApJ...473..322T}. As a side effect of the
magnetic dissipation, a much higher pair multiplicity is expected compared to
ordinary pulsars. \citet{2021ApJ...922L...7B} estimated the pair density in the
magnetospheres of magnetars using their typical X-ray luminosity:
\begin{equation}
    \label{eq:n-pair}
    n_{\pm}(r) \sim 10^{10}\,r_{9}^{-3} \left( \frac{L_\mathrm{keV}}{10^{35}\,\mathrm{erg/s}}\right)\,\mathrm{cm^{-3}}.
\end{equation}
We can use this to estimate the gyrofrequency and plasma frequency at the nonlinear radius $r_\mathrm{NL}$:
\begin{align}
    \label{eq:omega-p}
    \omega_B(r_\mathrm{NL}) &\sim 8.0\times 10^{14}\,B_{0,14}^{-1/2}L_\mathrm{wave,42}^{3/4}\,\mathrm{Hz}, \\
    \omega_{p}(r_\mathrm{NL}) &\sim 1.2\times 10^{11}\, B_{0,14}^{-3/4}L_\mathrm{wave, 42}^{3/8}L_\mathrm{keV,35}^{1/2}\,\mathrm{Hz}.
\end{align}
Both frequencies are much higher than the typical GHz frequency seen in FRBs and
satisfy $\omega_B \gg \omega_p \gg \omega$, hence we expect that the plasma
should remain in the MHD regime and exhibit collective behavior in response to
potential $E > B$ regions in the wave.

\begin{figure*}[t]
    \centering
    \begin{minipage}{0.54\textwidth}
        \includegraphics[width=\textwidth]{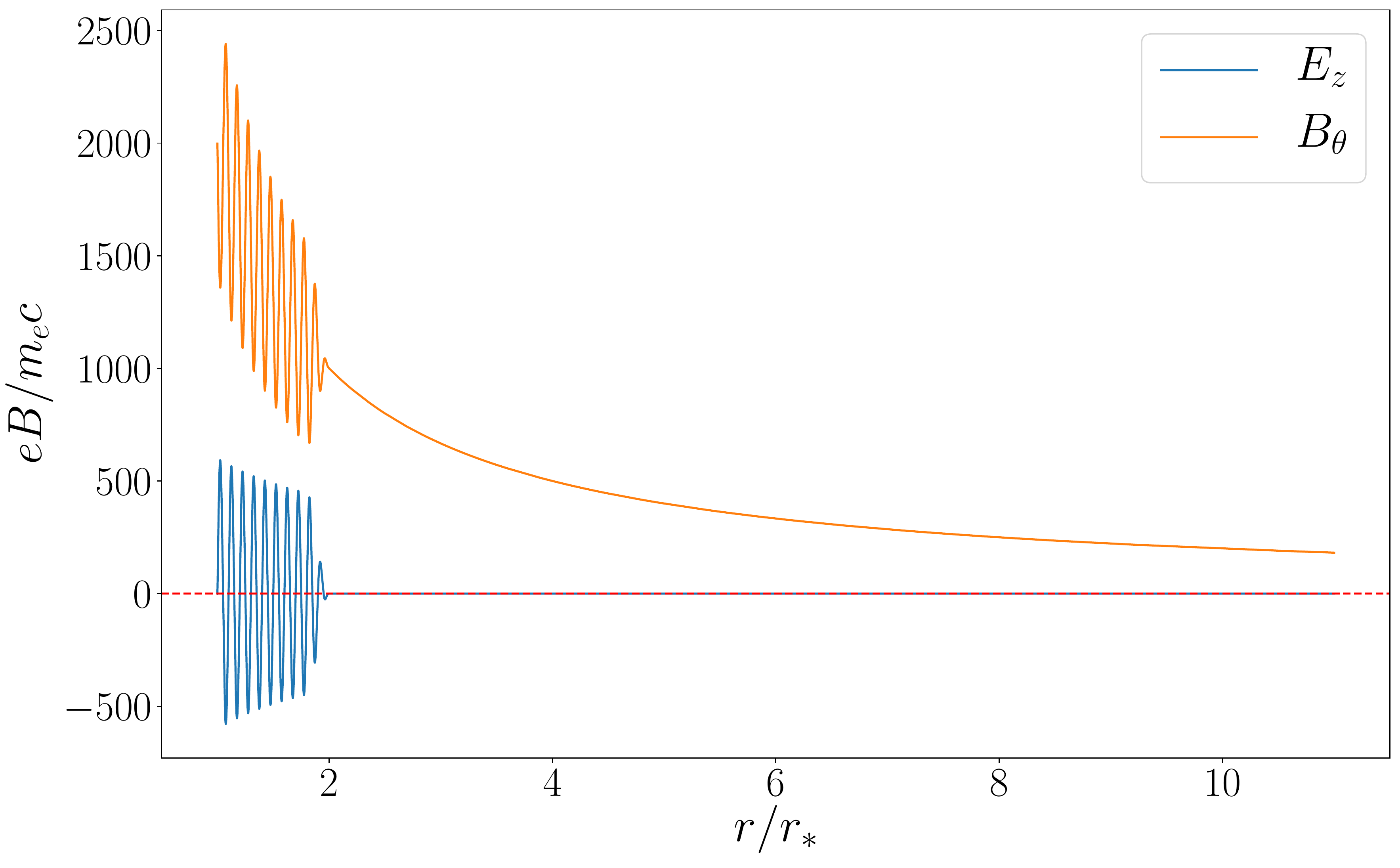}
    \end{minipage}
    \begin{minipage}{0.44\textwidth}
        \includegraphics[width=\textwidth]{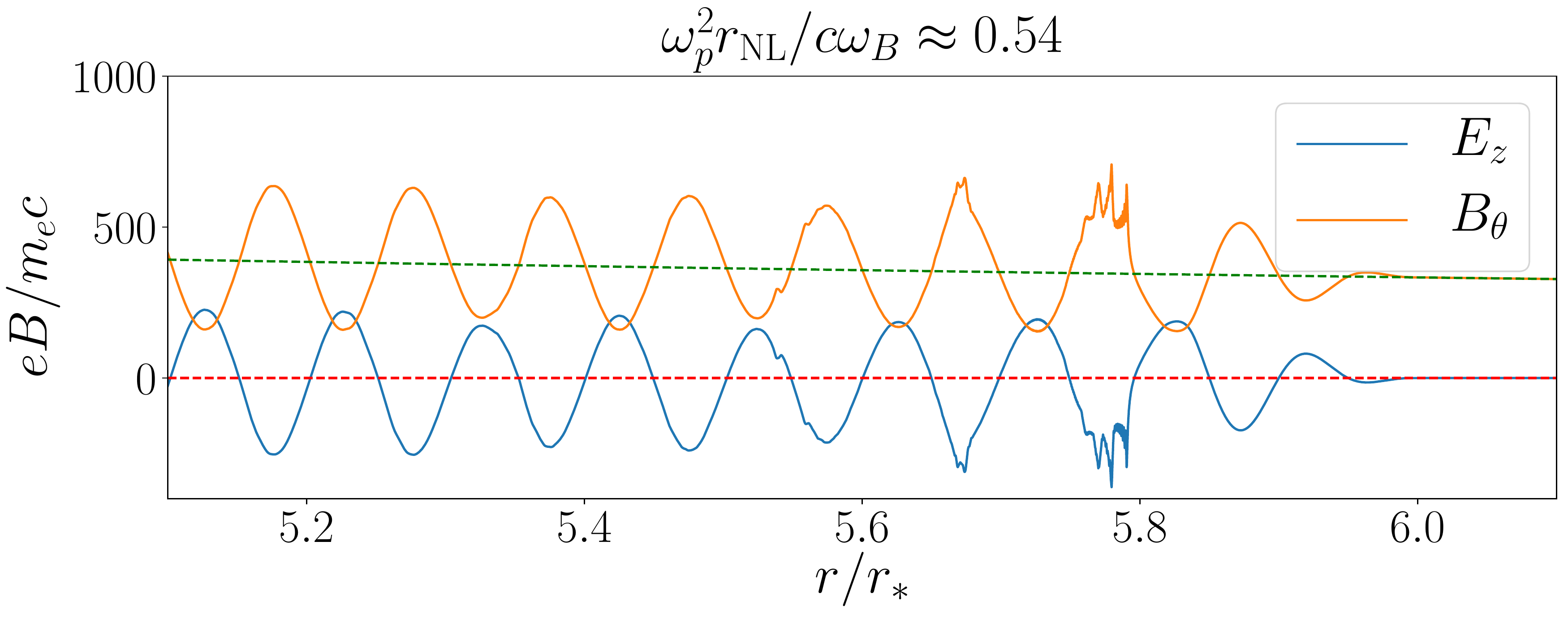}
        \includegraphics[width=\textwidth]{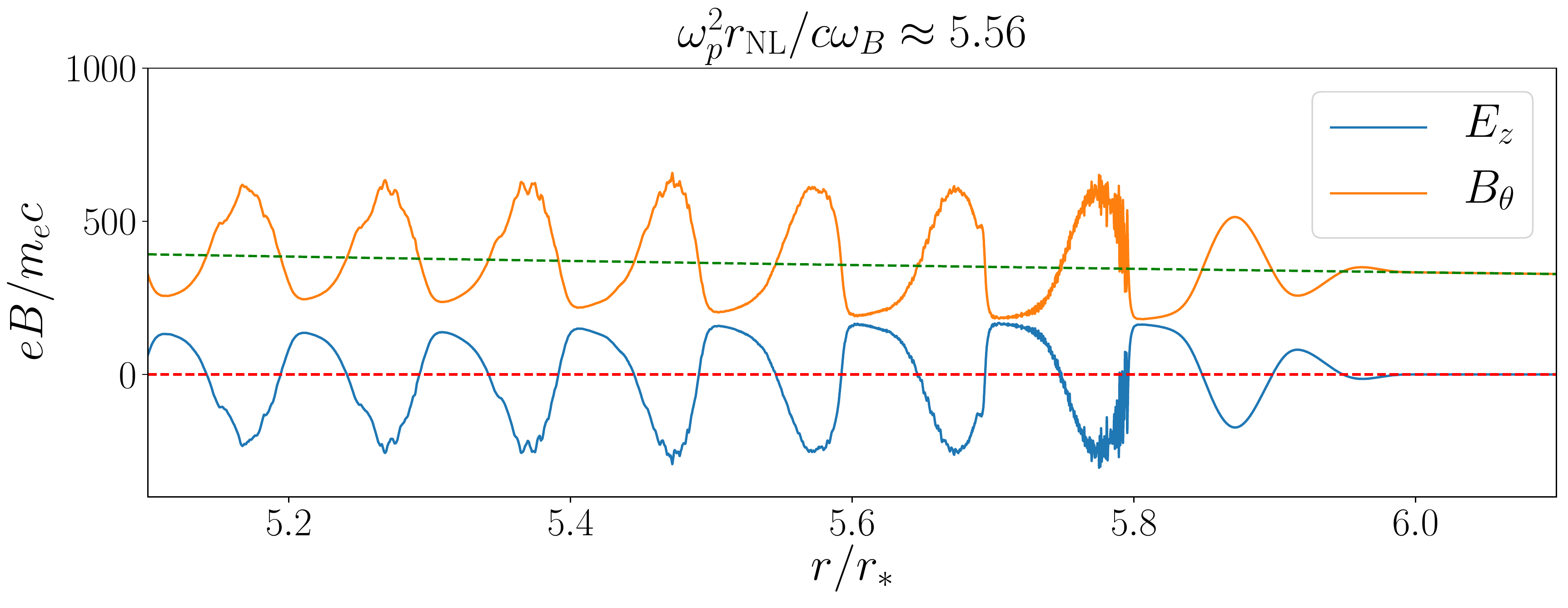}
    \end{minipage}
    \vspace{-0.3cm}
    \caption{\footnotesize \emph{Left}: Snapshot of the early stage of a
      simulation at $t \approx r_{*}/c$, where a wave is launched from $r = r_{*}$,
      before it becomes nonlinear near $r_\mathrm{NL} \approx 2.7r_{*}$.
      \emph{Right}: Snapshot of two simulations at $t \approx 5r_{*}/c$. The only
      difference between the two runs is $\omega_{p}$. The bottom run clearly
      shows the development of a shock in the leading wavelength, while the top
      run develops $E > B$ regions. Dashed lines in the right panel show the background fields $B_\mathrm{bg}$ (green) and $E_\mathrm{bg}$ (red).}
    \label{fig:1dsim}
\end{figure*}

The propagation of a fast wave in a constant background magnetic field where
$\delta B/B_\mathrm{bg}$ is close to 1/2 was studied by
\citet{2003MNRAS.339..765L}. The wave eventually steepens into a shock due to
nonlinearity, and he calculated the time scale for shock formation in the MHD
framework. For intermediate wave amplitudes $\delta B < B_\mathrm{bg}/2$, the
shock formation time is $t\sim \sigma/\omega$, where $\sigma$ is the plasma
magnetization and $\omega$ is the wave frequency, while this time decreases to
zero when the wave amplitude approaches $B_\mathrm{bg}/2$.

In the case of a decreasing background magnetic field however, this calculation
may not apply directly. If $B_\mathrm{bg}$ varies quickly enough, the wave may
not have enough time to steepen, and $E > B$ regions may be forced to develop.
As a result, plasma may be accelerated in these regions and become demagnetized,
at which point MHD assumptions no longer hold. On the other hand, if
$B_\mathrm{bg}$ varies relatively slowly, but faster than the shock formation
time at constant background field, then the wave may steepen much faster due to
$\delta B/B_\mathrm{bg}$ quickly approaching $1/2$, reducing the timescale for
shock development. In the following sections, we will use direct PIC simulations
to study the evolution of a fms wave traveling across a decreasing background
magnetic field. Our goal is to derive a more appropriate criterion for shock
development, and estimate the dissipation power resulting from this process.

\section{Numerical Simulations}\label{sec:sims}

\subsection{Setup}\label{sec:setup}

To simplify the problem into a quasi-1D one, we substitute the magnetosphere of
a neutron star with that of an infinitely long cylinder of radius $r_{*}$
carrying a uniform current along $z$ direction. It is surrounded by a toroidal
magnetic field $B_{\theta,\mathrm{bg}} = B_{0}(r/r_{*})^{-1}$. We launch a fast
wave from the surface of the cylinder with initial amplitude
$\delta B_{0} \ll B_{0}$ and polarization $\delta E$ along $z$, such that the
wave propagates outwards along the radial direction $\hat{\boldsymbol{r}}$, and
the wave magnetic field $\delta B$ is along the background magnetic field. As
the wave propagates, its amplitude decreases as $\delta B\sim r^{-1/2}$ but its
relative amplitude increases as $\delta B/B_\mathrm{bg} \sim r^{1/2}$. This
configuration is qualitatively similar to the propagation of a spherical wave in
the magnetosphere of a neutron star, with the exception that
$\delta B/B_\mathrm{bg}$ increases much more slowly (as $r^{1/2}$ instead of
$r^{2}$). This modification allows the magnetization
$\sigma = B^{2}/4\pi n_{\pm}m_{e}c^{2}$ to decrease much slower with radius,
enabling us to cover a larger range of radii $r$ in a single numerical
simulation where $\sigma \gg 1$.

We carry out the simulations in 2D polar coordinates using our open source
GPU-accelerated PIC code
\emph{Aperture}\footnote{\url{https://fizban007.github.com/Aperture4.git}}. We
are primarily interested in the radial evolution of the wave, therefore we
simulate only a thin wedge in the $\theta$ direction, employing periodic
boundary condition on both sides of the wedge. Since the width of the wedge is
much smaller than a wavelength, this setup is quasi-1D, without resolving the
physics in the transverse direction. The outer radial boundary at
$r_\mathrm{out}$ is set to be open and allows the wave to freely escape. In
practice, we usually terminate the simulation before the wave reaches the outer
boundary, therefore the outer boundary condition does not affect the results
presented below. The box is initially filled with a cold $e^{\pm}$ pair plasma,
with $\omega_{B}\gg \omega_{p}$, where $\omega_{B} = eB/m_{e}c$ is the electron
cyclotron frequency and $\omega_{p} = \sqrt{4\pi n_{e}e^{2}/m_e}$ is the plasma
frequency. This condition can also be written as $\sigma \gg 1$, where
$\sigma = B^{2}/(4\pi n m_{e}c^{2}) = (\omega_{B}/\omega_{p})^{2}$ is the cold
plasma magnetization. The initial temperature of the plasma is
$\Theta = kT/m_{e}c^{2}\sim 10^{-2}$, therefore does not contribute
significantly to the plasma enthalpy. In order to resolve the gyroradius
$\rho_{e} = m_{e}c^{2}/(eB_{0})$ of the background magnetic field, we use a
uniform grid with $196\text{,}608$ cells in $r$ and $16$ cells in $\theta$. The
resolution allows us to maintain $\sigma \gtrsim 10^{3}$ throughout the
simulation domain. We use at least $100$ particles per cell per species in our
production runs. The simulations were carried out on the OLCF supercomputer
\emph{Summit}.

The pair density in the box is chosen to scale as $n_{\pm} \propto r^{-1}$. As a
result, $\omega_{p}^{2}/\omega_{B}$ is a constant throughout the simulation
domain. This is motivated by the magnetosphere of a neutron star, where
$n_{\pm}\sim r^{-3}$ similar to the dipole magnetic field, assuming the pair
multiplicity does not vary appreciably in the magnetosphere. In that case,
$\omega_{p}^{2}/\omega_{B}$ is also a constant throughout the simulation domain.
The plasma is initialized with a Maxwellian distribution. The Debye length of
this background plasma is larger than the typical gyroradius, thus it is well
resolved. Throughout the simulation domain, the magnetization scales as
$\sigma \sim r^{-1}$, and the plasma frequency scales as
$\omega_{p} \sim r^{-1/2}$. We ensure that $\omega_p > \omega$ in the whole computational domain, even at the outer edge.

\subsection{Results}
\label{sec:results}

\begin{figure}[t]
    \centering
    \includegraphics[width=0.48\textwidth]{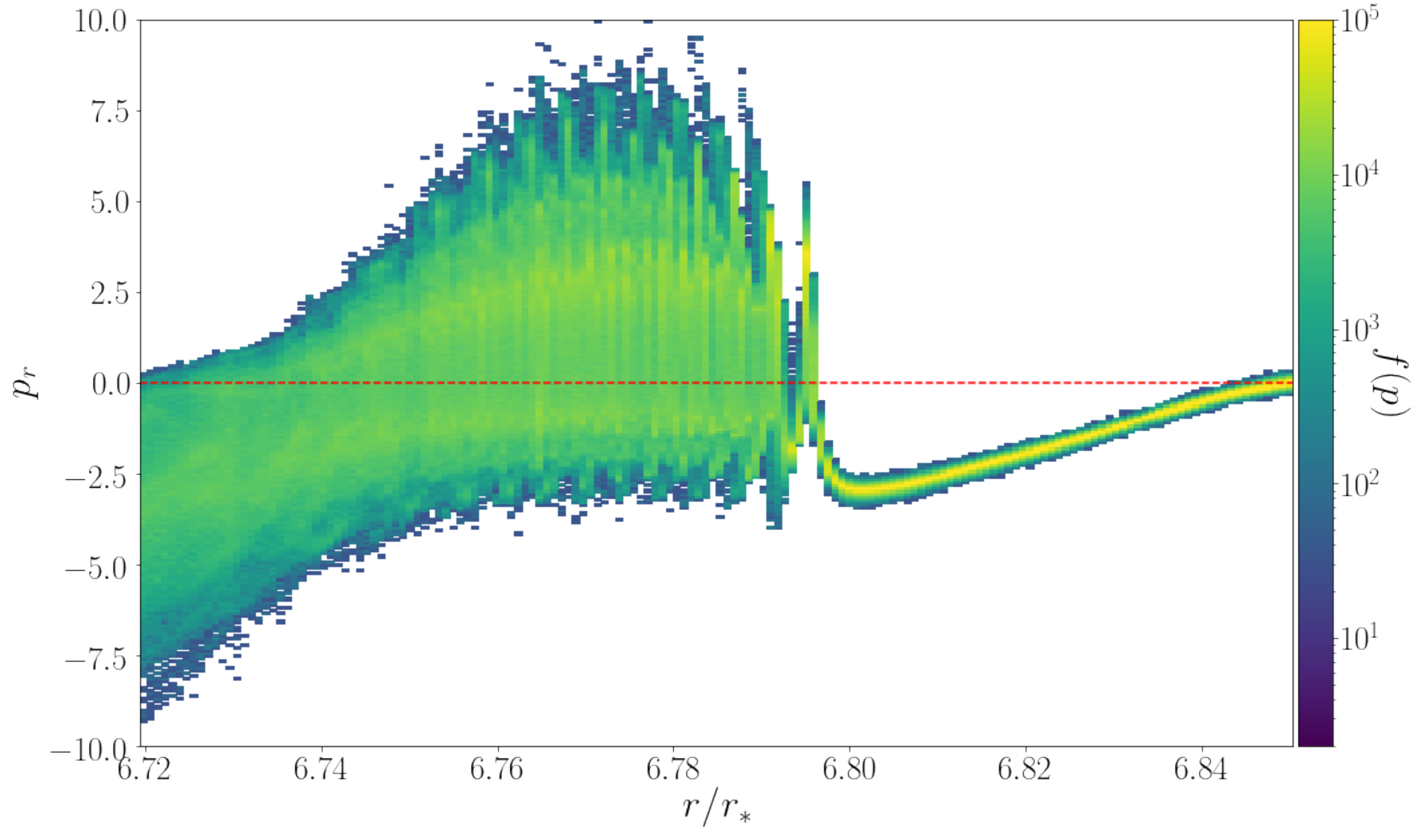}
    \caption{Phase space structure of the first shock in the wave train in the
      lab frame. The upstream is cold and drifts towards the shock front due to
      negative $(\boldsymbol{E}\times\boldsymbol{B})_{r}$. The phase space
      stream forms a solitonic structure as it encounters the shock and
      oscillates for a few cycles, eventually crossing itself and thermalizing
      to a temperature $kT_{2} \sim \gamma_{1}m_{e}c^{2}$. This hot plasma
      proceeds to become the upstream of the subsequent shock.}
    \label{fig:phase-space}
\end{figure}

We performed a series of numerical simulations which can be broadly categorized
into two qualitatively different regimes: either the wave nonlinearly steepens
into a shock, or $E > B$ regions develop and the wave does not steepen
appreciably. Figure~\ref{fig:1dsim} shows the result of two simulations with the
same parameters except $\omega_{p}$. For a larger plasma frequency, the waveform
deforms to avoid $E > B$ regions (see lower right panel of
Figure~\ref{fig:1dsim}). As a result, the corresponding peak shifts and
flattens, leading to a shock that forms behind the deformed wave. In the
upstream of this shock, the plasma drifts with $v_{r} < 0$ into the shock at
mildly relativistic speeds, and starts to gyrate in the higher magnetic field in
the downstream. In this region ahead of the shock, the plasma drift velocity in
the lab frame is always opposite to the wave propagation direction since the
total magnetic field $B$ is dominated by the background field $B_\mathrm{bg}$,
which has opposite sign as the wave magnetic field $\delta B$. This stream
initially forms a solitonic structure in the phase space, similar to what was
originally described by \citet{1988PhFl...31..839A} and commonly seen in numerical simulations of
perpendicular collisionless shocks
\citep[e.g.][]{1992ApJ...391...73G,2019MNRAS.485.3816P}. Eventually the stream
self-crosses in the phase space and becomes thermalized, with temperature $kT$
comparable to the drift kinetic energy of the upstream $\gamma_{1}m_{e}c^{2}$ (Figure~\ref{fig:phase-space}).

\begin{figure}[b]
    \centering
    \includegraphics[width=0.48\textwidth]{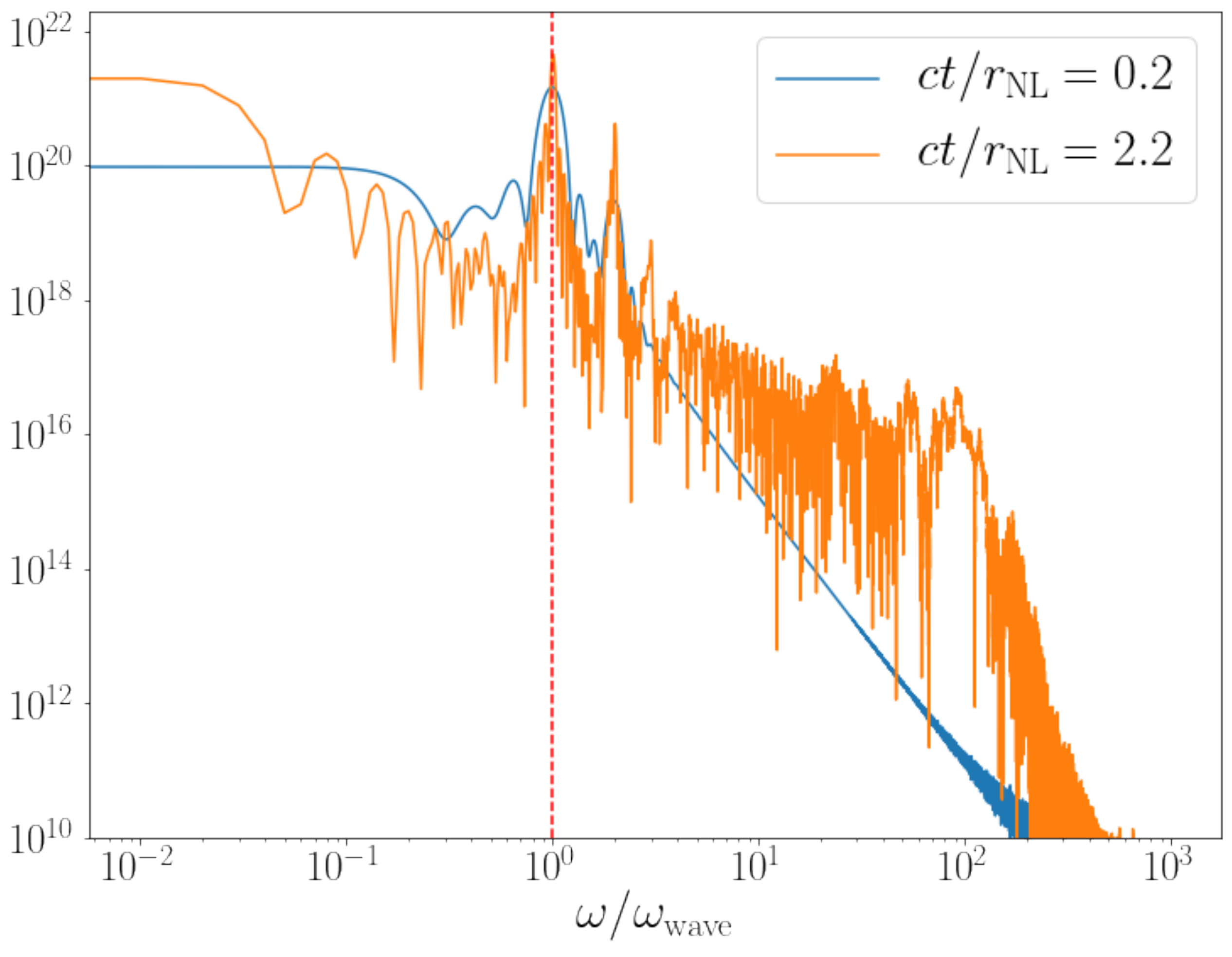}
    \caption{Power spectrum of the electric field before and after the wave
      passes through the nonlinear radius $r_\mathrm{NL}$ in the high
      $\omega_{p}$ case. Red dashed vertical line denotes the frequency of the
      fms wave. A fraction of the wave energy is converted to high frequency
      components close to the electron gyrofrequency. The data is taken from the
      run depicted in the lower right panel of Figure~\ref{fig:1dsim}.}
    \label{fig:spectrum}
\end{figure}

Contrary to \citet{2019MNRAS.485.3816P}, we do not observe a notable amount of
percursor waves generated at the shock front, even at the first shock where the
upstream plasma is cold. We believe this is due to the high magnetization values
in our simulations ($\sigma \gtrsim 10^{3}$), and the efficiency of generating the
precursor wave scales as $\sigma^{-1}$
\citep{2019MNRAS.485.3816P,2021PhRvL.127c5101S}. On the other hand, we do see a
significant amount of plasma oscillations post-shock, which is evident in the
right panels of Figure~\ref{fig:1dsim}. The power spectrum of the wave before
and after it passes through the nonlinear radius is shown in
Figure~\ref{fig:spectrum}. A fraction of the wave energy is deposited to higher
frequency modes as a power law, up to the electron gyrofrequency.

On the other hand, for a smaller $\omega_{p}$, the inertia of the plasma is not
sufficient to alter the waveform, leading to $E$ growing larger than $B$ (see
upper right panel of Figure~\ref{fig:1dsim}). The $E > B$ region act as a linear
accelerator, accelerating particles to much higher Lorentz factors than the
other case. As a result, the plasma becomes only marginally magnetized, with
$\omega_{B}/\gamma \gtrsim \omega$. This prevents subsequent wavelengths to
develop a shock as well. The later $E > B$ regions further accelerate the
particles. In Section~\ref{sec:shock-crit}, we will discuss semi-analytically
the criterion for the two different regimes and compare with the results
presented here.

\section{Criterion for Shock Development}
\label{sec:shock-crit}

We aim to derive a criterion for the two qualitatively different regimes
presented in Section~\ref{sec:results}. The change of wave electric field is the
result of a strong plasma current that is perpendicular to the background
magnetic field. In non-relativistic MHD this current takes the general form:
\begin{equation}
    \label{eq:j-perp}
    \boldsymbol{j}_\perp = \frac{\boldsymbol{B}\times\nabla P}{B^2} + \rho_e \frac{\boldsymbol{E}\times\boldsymbol{B}}{B^2} + \rho \frac{\boldsymbol{B}}{B^2}\times\frac{d\boldsymbol{v}}{dt},
\end{equation}
where $\boldsymbol{v}$ is the fluid bulk velocity, $P$ is pressure, $\rho_e$ is
charge density, and $\rho$ is the mass density. The first term is subdominant in
the high $\sigma$ limit, and the second term is perpendicular to the wave
electric field. The third term proportional to $d\boldsymbol{v}/dt$ is the
polarization current, which accounts for the inertia of the plasma. In the
relativistic regime, this generalizes to
\begin{align}
    \label{eq:j-perp-relativistic}
    \boldsymbol{j}_\perp &= \frac{\boldsymbol{B}\times\nabla P}{B^2} + \rho_e \frac{\boldsymbol{E}\times\boldsymbol{B}}{B^2} + (\rho_0 + u + P)\gamma^2 \frac{\boldsymbol{B}}{B^2}\times\frac{d\boldsymbol{v}}{dt}\nonumber\\
    &+\frac{\boldsymbol{B}}{B^2}\times\boldsymbol{v}\left(\frac{\partial P}{\partial t}+\boldsymbol{j}\cdot\boldsymbol{E}\right),
\end{align}
where $\rho_0$ and $u$ are the rest mass energy density and thermal energy
density in the fluid comoving frame, respectively, $\gamma$ is the bulk Lorentz
factor, and $d/dt=\partial/\partial t+\boldsymbol{v}\cdot\nabla$. When the fluid
Lorentz factor becomes significant, $\boldsymbol{j}_{\perp}$ is dominated by
the polarization current proportional to $\gamma^{2}$, which is along the
$\boldsymbol{E}$ field direction. This is the current that changes the wave form
and ultimately causes the shock to develop.

Consider the cylindrical wave in the numerical simulations of
Section~\ref{sec:results}. The perpendicular current is along the $z$ direction.
For a small amplitude fms wave in the high magnetization regime, this current is
typically very small compared to the displacement current,
$j_{z}/(\nabla\times B) \sim 1/\sigma$, and it vanishes in the force-free limit
since the rest mass energy density goes to zero. However, as
$\delta B/B_\mathrm{bg}$ approaches $1/2$, this current increases significantly,
up to $j_{z}\sim 0.1n_{\pm}ec$ as we measured empirically in our simulations.
This is the maximum rate at which the wave field $\delta E$ can change compared
to its force-free counterpart.

On the other hand, the background magnetic field changes on a completely
independent time scale which is set by the global structure of the
magnetosphere. The magnetic field of an infinite cylinder drops as
$B_\mathrm{bg}\sim r^{-1}$, while the wave field $\delta E$ naturally drops as
$\delta E \sim r^{-1/2}$. In order to avoid $E > B$ regions in the wave, the plasma
current $j_{\perp}$ needs to reduce the wave field such that it decreases at
least as fast as the background field. Consider the wave propagating from some
radius $r_{0}$ to $r_{0} + \Delta r$ during time $\Delta t \approx \Delta r/c$,
the criterion for avoiding $E > B$ becomes:
\begin{equation}
    \begin{split}
    4\pi j_{\perp}\Delta t &\gtrsim \delta E \left[\left(1 + \frac{\Delta r}{r_{0}}\right)^{-1/2} - \left(1 + \frac{\Delta r}{r_{0}}\right)^{-1}\right] \\
      & \approx \delta E \frac{1}{2}\frac{\Delta r}{r_{0}}.
    \end{split}
\end{equation}
If we evaluate this equation near the nonlinear radius, where $\delta E = \delta B \sim B_\mathrm{bg}/2$, and use $j_{\perp} \sim 0.1n_{\pm}ec$, we can write down the criterion for forming a shock at $r_\mathrm{NL}$:
\begin{equation}
    0.1\omega_{p}^{2}\gtrsim \frac{1}{2}\frac{c}{r_\mathrm{NL}}\frac{\omega_{B}}{2},
\end{equation}
where $\omega_{B} = eB_\mathrm{bg}/m_{e}c$ is the gyrofrequency in the
background magnetic field. Note that the coefficient of 1/2 depends on the geometry, and becomes 2 for spherical waves propagating in a dipole magnetic field. Therefore, a crude criterion for the formation of collisionless shocks at each wavelength when the fms wave comes near the nonlinear radius is:
\begin{equation}
    \label{eq:eta-threshold}
    \eta\equiv\frac{\omega_{p}^{2}r_\mathrm{NL}}{c\omega_{B}}\gtrsim
    \begin{cases}
      2.5,& \quad \text{if }\delta B/B_\mathrm{bg}\sim r^{1/2},\\
      10, & \quad \text{if }\delta B/B_\mathrm{bg}\sim r^{2}.
    \end{cases}
\end{equation}

The right panel of Figure~\ref{fig:1dsim} shows two contrasting simulations with
the dimensionless ratio $\eta$ below and above the threshold, and we see indeed
that in the latter case shocks are formed in the wave and no $E > B$ regions are
developed.

Taking this approximate criterion for development of shocks, we are interested
in whether typical FRBs produced in the inner magnetosphere will develop shocks
near the nonlinear radius, or propagate as vacuum waves as suggested by
\citet{2021ApJ...922L...7B}. For typical FRB parameters estimated in
Section~\ref{sec:fms-waves}, this ratio $\eta$ can be evaluated as:
\begin{equation}
    \label{eq:frb-eta}
    \eta = \frac{\omega_{p}^{2}r_\mathrm{NL}}{c\omega_{B}} \sim 7.8\times 10^{4}\,B_{0,14}^{-1/2}L_{\mathrm{wave},42}^{-1/4}L_{\mathrm{keV},35}.
\end{equation}
This is much larger than the threshold value of ${\sim}10$ for a dipole
background field, therefore we expect typical cosmological FRBs produced in the
magnetospheres of magnetars to steepen into shocks easily when they become
nonlinear due to propagation.

Since this dimensionless ratio does not depend on the frequency of the fms wave (as long as $\omega_p \gg \omega$),
the same conclusion applies to low frequency kHz fast waves launched directly
from the the magnetar either due to star quakes, or due to nonlinear conversion
from \alfven{} waves \citep{2021ApJ...908..176Y}. A kHz fms wave with isotropic
equivalent luminosity in a wide luminosity range of $10^{36}\,\mathrm{erg/s}$ to
$10^{42}\,\mathrm{erg/s}$ will inevitably steepen into shocks within the
magnetosphere. This mechanism can serve as an alternative way to launch shocks
from the magnetosphere of a magnetar, without the need for a relativstic ejecta
such as proposed by \citet{2020ApJ...900L..21Y,2022ApJ...933..174Y}. Formation
of the shock may shift a fraction of the wave energy into high frequency
components, as shown in Section~\ref{sec:results}. Eventually, the shock may also
become an efficient emitter of coherent radio waves when it propagates to large
distances from the star through the synchrotron maser mechanism \citep{2019MNRAS.485.3816P,2021PhRvL.127c5101S}.
\medskip

\section{Wave Dissipation through Shocks}\label{sec:dissipation}

In the highly magnetized limit where collisionless perpendicular shocks do form, dissipation
of the fast wave is mainly mediated through these shocks. We seek to describe
this dissipation process and evaluate how quickly waves of different wavelengths
will lose their energy through this mechanism.

Plasma heating through perpendicular shocks has been well-studied theoretically
\citep[see e.g.][]{1992ApJ...391...73G,2006ApJ...653..325A}. A synchrotron maser
forms at the front of the shock, coherently reflecting particles from the
upstream and thermalizing their bulk motion
\citep{1991PhFlB...3..818H,1992ApJ...391...73G}. By writing down the shock jump
condition in the downstream rest frame, one can find the heating of the
downstream as a function of the upstream Lorentz factor  and
magnetization $\sigma$, and the adiabatic index $\Gamma$. For a cold upstream, \citet{1992ApJ...391...73G} found that for $e^{\pm}$ plasma in the high $\sigma$ limit:
\begin{equation}
    \label{eq:T2-cold-upstream}
    \frac{kT_{2}}{\gamma_{12}m_{e}c^{2}} = \frac{2(\Gamma - 1)}{4 - \Gamma} + \frac{2\Gamma(\Gamma - 1)(2 - \Gamma)}{(4 - \Gamma)^{3}}\frac{1}{\sigma} + O\left(\frac{1}{\sigma^{2}}\right),
\end{equation}
where $\gamma_{12}$ is the upstream bulk Lorentz factor measured in the
downstream frame. The plasma moves in the plane perpendicular to the background magnetic field, therefore it is appropriate to use an adiabatic index of $\Gamma \to 3/2$.

In the case of an fms wave train described in this paper, a relativistic
perpendicular shock may form at every wavelength. Only the first shock in the
wave train may encounter a cold upstream, whereas every subsequent shock will
encounter the heated downstream of the preceding shock. In order to predict the
total dissipation power and plasma heating, we need the generalization of
Equation~\eqref{eq:T2-cold-upstream} in the case of a finite upstream
temperature. For pedagogical reasons we include a derivation in Appendix A,
where we write down the shock jump condition in the shock rest frame, then use
it to derive a constraint on shock heating:
\begin{equation}
    \label{eq:T2-hot-upstream}
    \frac{T_{2}}{T_{1}} \lesssim 1.7\gamma_{1}\left(1 + \frac{0.32}{\sigma} + O\left(\sigma^{-2}\right)\right),
\end{equation}
where $T_{1}$ and $T_{2}$ are the upstream and downstream plasma temperatures
measured in the respective comoving frame, $\gamma_1$ is the upstream bulk
Lorentz factor measured in the lab frame, and $\sigma$ is the upstream
magnetization. Across every shock except the first, the downstream is heated
with respect to the upstream by a factor of ${\sim}\gamma_{1}$. In the case of a
constant $\gamma_{1}$, the plasma temperature increases \emph{exponentially}
across many consecutive shocks. In our simulations, the upstream Lorentz factor
$\gamma_{1}$ remains at most mildly relativistic, especially when $\eta$ is far
above the threshold [Equation~\eqref{eq:eta-threshold}] and the shock is easily
formed.

This exponential heating cannot continue indefinitely as the wave will soon
deplete its energy. Several physical mechanisms will kick in before the wave
energy is depleted. The shock heating will become inefficient when the plasma is
heated to a high enough temperature such that the gyroradius of the hot
electrons become comparable to the wavelength of the fms wave,
$kT/m_{e}c^{2} \sim \omega_{B}/\omega$. Near the nonlinear radius
$r_\mathrm{NL}$ this temperature will be:
\begin{equation}
    \Theta_\mathrm{demag}\sim \frac{\omega_{B}}{\omega} \sim 8.0\times 10^{5}\,B_{0,14}^{-1/2}L_\mathrm{wave,42}^{3/4}.
\end{equation}
At this temperature, the plasma becomes
essentially demagnetized and collisionless shocks will not form within a
wavelength anymore.

Another mechanism that will limit plasma heating is the radiative cooling of the
downstream through synchrotron emission. Within the magnetar magnetosphere where
the background magnetic field is high, synchrotron cooling can place a strong
limit on the temperature of the plasma. This limit can be estimated by equating
the heating rate with the synchrotron cooling rate:
\begin{equation}
    \frac{kT_{2} - kT_{1}}{\omega^{-1}} \sim \frac{4}{3}\left(\frac{kT_{2}}{m_{e}c^{2}}\right)^{2}\sigma_{T}cU_{B}.
\end{equation}
Using $T_{2}\sim 1.7\gamma_{1}T_{1}$ and assuming $\gamma_{1}\sim 2$, the
upstream temperature at which shock heating balances synchrotron cooling becomes:
\begin{equation}
    \label{eq:T-balance}
    \Theta_{1}\equiv \frac{kT_{1}}{m_{e}c^{2}} \sim \frac{1}{6.4 \ell_{\lambda}},
\end{equation}
where $\ell_{\lambda}$ can be understood as the magnetic compactness over an fms
wavelength: $\ell_{\lambda} \equiv \sigma_{T}U_{B}\lambda/m_{e}c^{2}$. At this
temperature, subsequent shocks passing through the plasma will dissipate at a steady rate:
\begin{equation}
    \label{eq:dissipation}
    \dot{U} \sim \omega k(T_{2} - T_{1})n_{\pm} \sim 0.37\frac{\omega n_{\pm} m_{e}c^{2}}{\ell_{\lambda}}.
\end{equation}
All the energy dissipated in the shock is radiated away through synchrotron
radiation.

We can estimate the dissipation time scale for an FRB-like fms wave propagating
in the closed magnetic field line zone of the magnetar. Taking the estimates for
$r_\mathrm{NL}$ in Section~\ref{sec:fms-waves}, the magnetic compactness
$\ell_{\lambda}$ over the wavelength of a GHz FRB is:
\begin{equation}
    \ell_{\lambda} = 2.0\times 10^{-3}\left(\frac{r}{r_\mathrm{NL}}\right)^{-6}\,B_{0,14}^{-1}L_\mathrm{wave,42}^{3/2}.
\end{equation}
The compactness decreases rapidly with increasing radius. As a result, the
synchrotron balanced temperature is $\Theta_{1}\sim 10^{2}$ near the nonlinear
radius, and grows as $r^{6}$. The dissipation rate $\dot{U}$ is proportional to
$n_\pm/\ell_\lambda$, which scales as $r^3$. Due to the nature of exponential
heating over consecutive shocks, $\Theta_1 \sim 10^2$ can be reached very
quickly after only a few wavelengths. A typical $1\,\mathrm{ms}$ FRB contains
$10^6$ wavelengths, thus all but the first few wavelengths will pass through a
hot plasma at this radiation-balanced temperature.

Near the nonlinear radius, the total dissipation
power of an FRB assuming isotropic emission is:
\begin{equation}
    L_\mathrm{diss} \sim 4\pi r_\mathrm{NL}^{2}c\Delta t\dot{U} \sim 4.4\times 10^{42}\,\mathrm{erg/s},
\end{equation}
where $\Delta t$ is the duration of the burst.
This high dissipation power means that a significant fraction of the FRB energy
will be converted to radiation within the duration of the FRB itself. The
resulting radiation will have an energy of a few keV, and the $\Theta \sim 100$
plasma upscattering this soft X-ray will produce ${\sim}10\,\mathrm{MeV}$ soft
$\gamma$-rays which can convert into $e^{\pm}$ pairs. The total magnetic
compactness of the FRB wave train can be very high, $\ell_{B}\gtrsim 10^{3}$, and we expect
significant pair loading at the nonlinear radius where the FRB is dissipated. A
detailed calculation of the pair loading and subsequent radiation signature is
beyond the scope of the current paper, and will be studied in a future work.
Effectively, the FRB will likely not be able to escape the magnetosphere.

For a kHz fms wave launched near the magnetar surface, the wavelength is much larger and there are fewer shocks in total for the same duration. As a result, the magnetic compactness per wavelength $\ell_\lambda$ is significantly higher, $\ell_{\lambda} \sim 10^{3}$. Equation~\eqref{eq:T-balance} predicts a temperature much
lower than unity, suggesting that the plasma has plenty of time to cool to
nonrelativistic temperatures before encountering a second shock. It is therefore
a good approximation to treat each shock as propagating into a cold upstream,
and we expect only mild heating of the plasma from these shocks in the high
$\sigma$ region in the magnetar magnetosphere, instead of the exponential growth
of the plasma temperature in the case with no cooling. The exact heating rate is
given by Equation~\eqref{eq:T2-cold-upstream} and depends critically on the
upstream Lorentz factor $\gamma_1$. More study is required to quantify its dependence on
the plasma properties such as magnetization $\sigma$, gradient of the background
magnetic field, and plasma frequency $\omega_{p}$. Due to high magnetic
compactness in this region, synchrotron radiation
may interact nontrivially with the plasma in the shock. This topic is again out
of the scope of the current paper, but is worth looking into analytically or
through PIC simulations in the future.

\section{Discussions}

We have studied the propagation of a strong fast magnetosonic (fms) wave across
a slowly decreasing background magnetic field. The wave may become nonlinear, in
other words
$|\delta \boldsymbol{E}| \gtrsim |\delta \boldsymbol{B} + \boldsymbol{B}_\mathrm{bg}|$,
as a result of propagation. We found that when the plasma is well magnetized, its
collective response is important when the wave becomes nonlinear, and
collisionless perpendicular shocks may self-consistently develop at every
wavelength as a result. We derived a simple analytic criterion
[Equation~\eqref{eq:eta-threshold}] for when shocks will develop, and concluded
that these shocks will form for most cosmological FRBs as well as low frequency
kHz fast waves from the star. As a result of these shocks, FRBs produced deep
within the magnetosphere propagating in the closed field line zone may dissipate
completely within $1\,\mathrm{ms}$ when it becomes nonlinear, converting its
energy into $e^{\pm}$ pairs and high energy radiation. Low frequency kHz fast
waves do not suffer as much dissipation, but may form strong perpendicular
shocks that can propagate to large distances and produce FRBs through the
synchrotron maser mechanism.

One crucial simplification of the simulations presented in
Section~\ref{sec:sims} was that we considered the global magnetic field of a
current-carrying cylinder rather than a spherical star with dipole magnetic
field. As a result, we have $\delta B/B_\mathrm{bg} \sim r^{1/2}$ instead of the
realistic case where $\delta B/B_\mathrm{bg} \sim r^{2}$. Most notably, the
relative strength of the wave grows slower at large radii in the cylindrical
case, while it grows faster and faster in a dipole magnetosphere. This
difference may have implications on the long term behavior of the shocks formed
near $r_\mathrm{NL}$, and should be studied in the future using global PIC
simulations in spherical coordinates.

Despite our usage of PIC simulations to study this problem, the magnetized
regime where shocks form can be well-described by relativistic MHD.\@ An analytic
model based on MHD equations, similar to what was developed by
\citet{2003MNRAS.339..765L}, can give better estimates on the upstream Lorentz
factors $\gamma_{1}$ at each shock. Such a calculation will improve the
estimates of shock heating and provide better constraints on the wave
dissipation rate. More systematic PIC simulations of the shocks at a variety of
parameter regimes will also allow us to better understand the transition from
the well-magnetized MHD regime to the vacuum-like regime, and potentially test
the wave scattering theory proposed by \citet{2022PhRvL.128y5003B}.

Radiative effects and pair production at the shocks are also interesting
by-products in these low frequency fast waves when
they become nonlinear. As estimated in Section~\ref{sec:dissipation}, an FRB
dissipated through this mechanism may lead to strong pair loading and an X-ray
flare. Future works in this direction can potentially lead to better theoretical
understanding of the rich X-ray phenomenology concerning magnetars.

\begin{acknowledgments}

    We thank Yuri Levin, Andrei Beloborodov, Dmitri Uzdensky, and Greg Werner
    for helpful discussions. AC acknowledges support from Fermi Guest
    Investigation grant 80NSSC21K2027. AC and YY acknowledge support from NSF grant
    DMS-2235457. XL is supported by NSERC, funding reference \#CITA 490888-16 and the Jeffrey L. Bishop Fellowship. This work was also facilitated by Multimessenger Plasma Physics Center (MPPC), NSF grant PHY-2206608. This research used resources of the Oak
    Ridge Leadership Computing Facility at the Oak Ridge National Laboratory,
    which is supported by the Office of Science of the U.S. Department of Energy
    under Contract No. DE-AC05-00OR22725.

\end{acknowledgments}

\appendix
\section{Shock Jump Condition for A Magnetized Relativistic Fluid}\label{sec:app_shock}
For a magnetized relativistic fluid, its stress-energy tensor is given by
\begin{equation}
    T^{\mu\nu} = \left(\rho h + \frac{b^2}{4\pi}\right)u^\mu u^\nu + \left(P+\frac{b^2}{8\pi}\right)g^{\mu\nu} - \frac{b^\mu b^\nu}{4\pi}.
\end{equation}
Here $g$ is metric tensor, $h$ is specific enthalpy, $P$ is pressure, $u$ is 4-velocity, and $b^\mu=u_\nu F^{*\mu\nu}$ is magnetic field in the comoving frame, $b^2=b_\mu b^\nu$. All thermodynamical quantities are defined in the comoving frame of the fluid. We consider the fluid as an ideal gas consisting of collisionless particle with equation of state $P=\rho^\Gamma$ and specific internal energy dominated by relativistic thermal motion. Therefore $h=\kappa P/\rho$ is proportional to the fluid temperature $T$ and $\kappa=\Gamma/(\Gamma-1)$.

The fluid evolves under the conservation laws of matter and stress-energy, plus the Maxwell equations.
\begin{eqnarray}
\nabla_\mu(\rho u^\mu)&=&0\\
\nabla_\mu T^{\mu\nu}&=&0\\
\nabla_\mu F^{*\mu\nu}&=&0.
\end{eqnarray}
For the shock in 1D with perpendicular magnetic field,  we can reach the following shock jump condition in the shock frame
\begin{eqnarray}
\llbracket\rho\gamma\beta\rrbracket &=& 0 \label{eqn:shock1}\\
\left\llbracket b\gamma\beta\right\rrbracket &=& 0\label{eqn:shock2}\\
\left\llbracket(\rho h + \frac{b^2}{4\pi})\gamma^2\beta\right\rrbracket &=& 0 \label{eqn:shock3}\\
\left\llbracket(\rho h + \frac{b^2}{4\pi})\gamma^2\beta^2 + \frac{b^2}{8\pi}+P\right\rrbracket &=& 0.\label{eqn:shock4}
\end{eqnarray}
Here we define the operator $\llbracket Q\rrbracket=Q_1-Q_2$, is the difference of quantity $Q$ across the shock with the subscript $1$ standing for upstream and $2$ for downstream.
Introducing the velocity ratio $r\equiv \beta_2/\beta_1$, we have
\begin{equation}
    \frac{\gamma_2}{\gamma_1} = \sqrt{\frac{1-\beta_1^2}{1-r^2\beta_1^2}}.
\end{equation}
\begin{equation}\label{eqn:shock5}
    \frac{\rho_2}{\rho_1} = \frac{b_2}{b_1}= \frac{\gamma_1\beta_1}{\gamma_2\beta_2} = \frac{1}{r}\sqrt{\frac{1-r^2\beta_1^2}{1-\beta_1^2}}.
\end{equation}
From Equation~\ref{eqn:shock3}, the energy ratio and enthalpy ratio across the shock is
\begin{equation}
    \frac{\rho_2 h_2 + b_2^2/4\pi}{\rho_1 h_1 + b_1^2/4\pi} = \frac{\gamma_1^2\beta_1}{\gamma_2^2\beta_2}
\end{equation}
\begin{equation}
    \frac{h_2}{h_1} = \frac{\gamma_1}{\gamma_2}+\sigma_1\frac{\gamma_1}{\gamma_2}\left(1-\frac{1}{r}\right)
\end{equation}
where $\sigma=b^2/4\pi \rho h$ is the hot plasma magnetization.

Using Equation~\ref{eqn:shock4} we obtain an equation for $r$
\begin{equation}
    (r-1)\left[ \left(1-\frac{1}{\kappa}\right)(1+\sigma_1)\beta_1^2 r^2 -\left(\frac{1}{\kappa}+\frac{\sigma_1}{2}\right)r+\sigma_1\left(\frac{1}{\kappa}-\frac{1}{2}\right) \right]=0
\end{equation}
which admits a nontrivial solution
\begin{equation}
    r = \frac{2+\kappa\sigma_1+\sqrt{(2+\kappa\sigma_1)^2+8\beta_1^2(\kappa-1)(\kappa-2)\sigma_1(\sigma_1+1)}}{2\beta_1^2(\kappa-1)(\sigma_1+1)}.
\end{equation}
For physical conditions where the shock compresses the fluid, $r<1$
\begin{equation}
    \beta_1>\sqrt{\frac{\Gamma-1+\sigma_1}{1+\sigma_1}}=\frac{\Gamma P+b^2/4\pi}{\rho h+b^2/4\pi}=\beta_F
\end{equation}
where $\beta_F$ is the ratio of the fast magnetosonic speed over the speed of light. The upstream fluid must move supersonically to trigger the shock.

The (comoving) temperature of the fluid is given by $T\propto P/\rho\propto h$, therefore the shock will heat the fluid by
\begin{equation}
    \frac{T_2}{T_1} = \frac{\gamma_1}{\gamma_2}\left(1+\sigma_1-\frac{\sigma_1}{r}\right).
\end{equation}
Assuming that the shock moves at the speed $\beta_\mathrm{sh}c$, we can express the temperature ratio in terms of the lab frame bulk Lorentz factors $\tilde\gamma_1$ and $\tilde\gamma_2$
\begin{equation}
    \frac{T_2}{T_1} = \frac{\tilde\gamma_1+\beta_\mathrm{sh}\sqrt{\tilde\gamma_1^2-1}}{\tilde\gamma_2-\beta_\mathrm{sh}\sqrt{\tilde\gamma_2^2-1}}\left(1+\sigma_1-\frac{\sigma_1}{r}\right).
\end{equation}
We specialize to a 2D relativistic plasma where $\Gamma = 3/2$ and $\kappa = 3$. In the limit of an ultra-relativistic fluid, $\beta_1\rightarrow 1$, and high upstream magnetization, $\sigma_1\rightarrow\infty$, we have
\begin{equation}
    r=1-\frac{2}{5\sigma_{1}}+\frac{44}{125\sigma_{1}^2}+\mathcal{O}(\sigma_{1}^{-3})
\end{equation}
and
\begin{equation}
    \frac{T_2}{T_1} = \frac{\tilde\gamma_1+\beta_\mathrm{sh}\sqrt{\tilde\gamma_1^2-1}}{\tilde\gamma_2-\beta_\mathrm{sh}\sqrt{\tilde\gamma_2^2-1}}\left(\frac{3}{5}+\frac{24}{125\sigma_1}+\mathcal{O}(\sigma_{1}^{-2})\right).
\end{equation}

In the context of perpendicular shocks launched by the nonlinear steepening of fast magnetosonic waves described in Section~\ref{sec:sims}, the values of $\tilde{\gamma}_{1}$ and $\tilde{\gamma}_{2}$ are given by the upstream and downstream electromagnetic fields. In the downstream, $B = B_\mathrm{bg} + \delta B$ where $\delta B \lesssim B_\mathrm{bg}/2$, and the fluid bulk velocity is simply given by the $\boldsymbol{E}\times\boldsymbol{B}$ drift: $\tilde{v}_{2}\sim \delta E/B \lesssim 1/3$ and $\tilde{\gamma}_{2} \lesssim 1.06$. Since $\beta_\mathrm{sh} < 1$ by construction, we can estimate the shock heating rate:
\begin{equation}
    \frac{T_{2}}{T_{1}} \lesssim 1.7\tilde{\gamma}_{1}\left(1 + \frac{0.32}{\sigma_1}+\mathcal{O}(\sigma_{1}^{-2})\right)
\end{equation}

\bibliographystyle{aasjournal}

\end{document}